\documentclass[conference]{IEEEtran}
\IEEEoverridecommandlockouts
\ifCLASSINFOpdf
   \usepackage[pdftex]{graphicx}
\else
\fi
\usepackage{amssymb,amsmath}

\usepackage[tight,footnotesize]{subfigure}
\begin{document}
%
\title{RNA structure characterization from chemical mapping experiments}

\author{Sharon~Aviran, Julius~B.~Lucks, and~Lior Pachter%
\thanks{Sharon Aviran is with the Center for Computational Biology and the  California Institute for Quantitative Biomedical Research (QB3), University of California, Berkeley, CA 94720, saviran@berkeley.edu. Julius B. Lucks is with the School of Chemical and Biomolecular Engineering, Cornell University, Ithaca, NY 14853, jblucks@cornell.edu. Lior Pachter is with the Departments of Mathematics, Molecular and Cell Biology, and Electrical Engineering and Computer Science, University of California, Berkeley, CA 94720, lpachter@math.berkeley.edu.}%
\thanks{This work was supported in part by a Celera Innovation
Fellowship to Sharon Aviran and by Tata Consultancy Services (TCS),
through grants to the Center for Computational Biology at the
University of California, Berkeley.}}


\maketitle

\begin{abstract}
Despite great interest in solving RNA secondary structures due to
their impact on function, it remains an open problem to determine
structure from sequence. Among experimental approaches, a promising
candidate is the ``chemical modification strategy'', which involves
application of chemicals to RNA that are sensitive to structure and
that result in modifications that can be assayed via sequencing
technologies. One approach that can reveal paired nucleotides via
chemical modification followed by sequencing is SHAPE, and it has been used in conjunction with capillary electrophoresis (SHAPE-CE) and high-throughput sequencing (SHAPE-Seq). The solution of mathematical inverse problems is needed
to relate the sequence data to the modified sites, and a number of approaches have been previously suggested for SHAPE-CE, and separately for SHAPE-Seq analysis.

Here we introduce a new model for inference of chemical modification experiments, whose formulation results in closed-form maximum likelihood estimates that can be easily applied to data. The model can be specialized to both SHAPE-CE and SHAPE-Seq, and therefore allows for a direct comparison of the two technologies. We then show that the extra information obtained with  SHAPE-Seq but not with SHAPE-CE is valuable with respect to ML estimation.  
\end{abstract}



%
\IEEEpeerreviewmaketitle

\section{Introduction}

\begin{figure}[t!]
\centering
\includegraphics[width=3.4in]{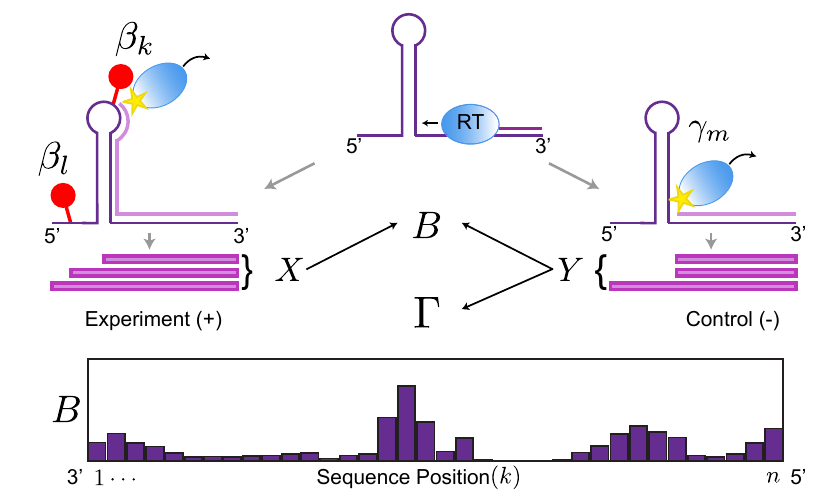}
\caption{Overview of a SHAPE-Seq chemical mapping experiment, model,
and statistical analysis.} \label{fig:SHAPE_Seq_Overview}
\end{figure}

RNA dynamics are increasingly recognized as central components of cellular function, controlling key processes such as gene regulation, antiviral defense, and environmental sensing~\cite{Sharp-RNA-Centrality-Review-2009, Chang-lncRNA-Review-2009,  Collins-RNA-Syn-Bio-Review-2006}. Strong links between RNA structure and function underlie the importance of structural analysis, which greatly benefits from the wealth of information provided by existing and emerging chemical mapping techniques~\cite{Weeks-ChemProbing-Review-2010}. In chemical mapping experiments, a chemical reagent modifies RNA molecules in a structure-dependent fashion. Depending on the reagent used, four distinct types of information can be gleaned, including spatial nucleotide contact information, solvent accessibility of the backbone, the local electrostatic environment adjacent to each nucleotide, and local nucleotide flexibility~\cite{Weeks-ChemProbing-Review-2010,
Rocca-Serra-Mapping-Exp-Standards-2011}. This information is then
used to infer RNA structural dynamics, either independently or in
conjunction with structure prediction algorithms~\cite{Mathews-2004, Weeks-Review-2010}. The modification location is
detected by means of conversion to cDNA using reverse transcriptase
(RT), whereby transcription is blocked at the sites of modification
(see illustration in Fig.~\ref{fig:SHAPE_Seq_Overview}). This
generates a pool of cDNA fragments that begin at the $3'$ end of the
molecule and terminate at the modified sites, or possibly at sites
where there was natural RT
dropoff~\cite{Weeks-Nature-Protocols-2006}. Traditionally, the cDNA
fragments have been resolved and quantified with capillary
electrophoresis (CE)~\cite{Weeks-Nature-Protocols-2006}, although
recently next-generation sequencing (NGS) technologies with much
higher throughput have been used
instead ~\cite{SHAPE-Seq-2011}.

There are several challenges in interpreting chemical mapping data
obtained from reverse transcription, irrespectively of the fragment
quantification method that follows it. Primarily, in molecules with
multiple modifications, only the first one (i.e., the closest to the
$3'$ end) is revealed (see Fig.~\ref{fig:SHAPE_Seq_Overview}), and
thus less information is available about the $5'$ region of the
molecule. Second, RT's natural propensity to terminate at any site
needs to be decoupled from modification-based termination, and this
effect is controlled for in a separate control experiment. Finally,
experimental variations need to be controlled for when combining
measurements. 

In previous work, we introduced a stochastic model for a specific
next-gen sequencing based chemical modification experiment called
SHAPE-Seq ({\bf s}elective $2'$-{\bf h}ydroxyl {\bf a}cylation
analyzed by {\bf p}rimer
{\bf e}xtension followed by {\bf seq}uencing)
~\cite{SHAPE-Seq-ML-2011}. Here we extend that work by presenting a
more general model that entails fewer assumptions. In addition
to capturing our previous SHAPE-Seq model as a special case, it is
suitable for other experimental protocols, such as SHAPE-CE (SHAPE
followed by capillary electrophoresis). The new model has the added
advantage that the generalization reveals a simplified maximum
likelihood estimation scheme that leads to an elegant and fast
approach for recovering chemical modification signal. Finally,  we show how the general framework can be used to directly compare the power of SHAPE-CE to SHAPE-Seq. A key result is that SHAPE-Seq improves on SHAPE-CE not only by allowing for multiplexing but also by measuring extra information that can be utilized in the statistical inference framework we propose.

\section{Modeling sequencing-based chemical mapping} \label{sec:general_model}

We consider an RNA molecule that contains $n$ sites, numbered $1$ to
$n$ according to their sequence-position with respect to the
molecule's $3'$ end (see Fig.~\ref{fig:SHAPE_Seq_Overview}), where
the $3'$ end is excluded from analysis and assumes position $0$. A
cDNA fragment of length $k$ that maps to the sequence between sites
$0$ and $k-1$ ($1\leq k \leq n$) is called a \emph{$k$-fragment},
and a full transcript of length $n+1$ is called a \emph{complete
fragment}. In a SHAPE experiment, often called the $(+)$ channel,
the RNA is treated with an electrophile that reacts with
conformationally flexible nucleotides to form $2'$-$O$-adducts. We
define the {\em relative reactivity} of a site, $\beta_k$, to be the
probability of adduct formation at that site. In the control
experiment, called the \emph{$(-)$ channel}, the primary source of
incomplete fragments is RT's natural dropoff while transcribing the
molecule. Because its propensity to drop may vary along sites, we
define the \emph{dropoff propensity at site $k$}, $\gamma_k$, to be
the conditional probability that transcription terminates at site
$k$, given that RT has reached this site. Therefore, associated with
the RNA molecule are $2n$ probabilities: $B = (\beta_1, \ldots,
\beta_n)$, $0\leq \beta_k \leq 1 \mbox{\ } \forall \mbox{\ }k$, and
$\Gamma=(\gamma_1, \ldots, \gamma_n)$, $0\leq \gamma_k \leq 1
\mbox{\ } \forall \mbox{\ }k$, which we wish to estimate from
sequencing data.

While we can readily infer the natural dropoff propensities from the
$(-)$ channel data alone~\cite{SHAPE-Seq-ML-2011}, the fragments
observed in the $(+)$ channel reflect the combined effects of
natural dropoff and chemical modification. A point that is key to
interpreting chemical mapping data is that a $k$-fragment is assumed
to be generated when site $k$ is the site that is \emph{first}
encountered by RT, regardless of the number of adducts that formed
upstream of $k$. Assuming that adduct formations at the various
nucleotides are statistically independent, the probability that a
molecule is modified at site $k$ (and possibly also at subsequent
sites) is
\begin{equation} \label{eq:first_adduct_at_k}
Prob\left(\mbox{first adduct at site $k$}\right) = \beta_k
\prod_{i=1}^{k-1}
(1-\beta_i)
\end{equation}
for all $1\leq k\leq n$. Incorporating the natural degradation in
the elongating pool of modified molecules, we have
\begin{equation} \label{eq:modification_based_prob}
Prob\left(\parbox{0.73in}{\centering $k$-fragment from
modification}\right) = \prod _{i=1}^{k-1} (1-\gamma_i) \times
\beta_k \prod_{i=1}^{k-1} (1-\beta_i).
\end{equation}
We assume all other fragments originate from natural dropoff, either
from unmodified or modified molecules, thus accounting for the
following probability:
\begin{eqnarray} \label{eq:structure_based_prob}
& & Prob\left(\mbox{$k$-fragment from natural dropoff}\right) \\
\nonumber & = & Prob\left(\parbox{0.47in}{\centering dropoff at site
$k$} \Big|\parbox{0.63in}{\centering no adduct at any site $l \leq
k$}\right) \times Prob\left(\parbox{0.63in}{\centering no adduct at
any site $l \leq k$}\right) \\ \nonumber & = & \gamma_k \prod
_{i=1}^{k-1} (1-\gamma_i) \prod_{i=1}^{k} (1-\beta_i).
\end{eqnarray}
Taken together, Eqs.~\ref{eq:modification_based_prob}
and~\ref{eq:structure_based_prob} imply
\begin{eqnarray} \label{eq:k_frag_prob}
& & Prob\left(\mbox{$k$-fragment in $(+)$ channel}\right) \\
\nonumber & = & \Big[ 1 - (1-\gamma_k)(1-\beta_k)\Big]
\prod_{i=1}^{k-1}
(1-\gamma_i)(1-\beta_i) 
\end{eqnarray}
for all $1\leq k\leq n$. Finally, because complete fragments can
only arise from natural dropoff, we have
\begin{eqnarray} \label{eq:complete_frag_prob}
& & Prob\left(\mbox{complete fragment in $(+)$ channel}\right) \\
\nonumber & = & \prod _{i=1}^{n} (1-\gamma_i) \prod_{i=1}^{n}
(1-\beta_i) = \prod _{i=1}^{n} (1-\gamma_i)(1-\beta_i).
\end{eqnarray}

Assuming we observe $(X_1, \ldots, X_{n+1})$ $k$-fragment and
complete-fragment counts in the $(+)$ channel, and similarly, $(Y_1,
\ldots, Y_{n+1})$ fragment counts in the $(-)$ channel, the
likelihood of observing the entire sequencing data is given by
\begin{eqnarray} \label{eq:likelihood}
\mathcal{L}(B, \Gamma)
& = & \prod_{k=1}^n \Big[\gamma_k \prod_{i=1}^{k-1}
(1-\gamma_i)\Big]^{Y_k} \\ \nonumber & & \prod_{k=1}^n \Big[\Big( 1
- (1-\gamma_k)(1-\beta_k)\Big)
\\ \nonumber & & \prod_{i=1}^{k-1} (1-\gamma_i)(1-\beta_i)\Big]^{X_k} \\ \nonumber &
& \Big[\prod_{i=1}^{n} (1-\gamma_i)\Big]^{Y_{n+1}} \Big[\prod
_{i=1}^{n} (1-\gamma_i)(1-\beta_i)\Big]^{X_{n+1}}.
\end{eqnarray}

\section{Maximum-likelihood estimation}

In this section, we use the likelihood formulation in
Eq.~\ref{eq:likelihood} to show that  
 \textbf{\emph{the ML estimates are given by}}


\begin{equation} \label{eq:Seq_ML_scheme}
\beta_k^{\ast} = \max \left\{0, \mbox{\ }
\frac{\frac{X_k}{\sum_{i=k}^{n+1} X_i} - \frac{Y_k}{\sum_{i=k}^{n+1}
Y_i}}{1 - \frac{Y_k}{\sum_{i=k}^{n+1} Y_i}}\right\}, \mbox{\ \ } 1\leq k \leq n.
\end{equation}

\vspace{0.25cm}

\noindent Moreover, as will become clear from the derivation below, the likelihood formulation in Eq.~\ref{eq:likelihood} and its optimization can be readily extended to accommodate data from multiple replicates. One can then estimate the reactivities from multiple sources of data simultaneously and in a straightforward manner, and without any further assumptions or estimation of the statistical inter-experiment variation.

We start by rearranging terms in the log-likelihood function and
writing it as the following sum of $n$ terms:
\begin{eqnarray} \label{eq:log_likelihood} \log\mathcal{L}(B,
\Gamma)
& = & \sum_{k=1}^{n} \Big[ \sum_{i=k+1}^{n+1} (X_i + Y_i)
\log (1-\gamma_k) \\ \nonumber & & + {} \sum_{i=k+1}^{n+1}X_i \log
(1-\beta_k) + Y_k \log \gamma_k \\ \nonumber & & + {} X_k \log\big(
1 - (1-\gamma_k)(1-\beta_k)\big) \Big].
\end{eqnarray}
Eq.~\ref{eq:log_likelihood} suggests that $\log\mathcal{L}(B,
\Gamma)$ is separable in the pairwise variables $(\beta_k,
\gamma_k)$, hence each of the $n$ two-dimensional functions can be
optimized separately. To simplify notation, we introduce the
constants $S_k=\sum_{i=k+1}^{n+1} (X_i + Y_i)$,
$R_k=\sum_{i=k+1}^{n+1} X_i$, and the functions
\begin{eqnarray} \label{eq:two_dim_func}
l_k(\beta_k, \gamma_k) & = & S_k \log (1-\gamma_k) + R_k \log
(1-\beta_k) \\ \nonumber & & + Y_k \log \gamma_k + X_k \log\big(
1 - (1-\gamma_k)(1-\beta_k)\big) 
\end{eqnarray}
for $1\leq k\leq n$, such that $\log\mathcal{L}(B,
\Gamma)=\sum_{k=1}^{n} l_k(\beta_k, \gamma_k)$.

We now optimize $l_k(\beta_k, \gamma_k)$ under the
assumption that all fragment counts are positive. 
In that case, $\gamma_k$ is bound to lie in $(0,1)$ and $\beta_k$
cannot exceed $1$, but the constraint $\beta_k \geq 0$ is not
inherent in the function and needs to be imposed during
optimization. If we relax it, we find the optimal solution by
setting the partial derivatives to zero, as follows
\begin{eqnarray} \label{eq:partial_deriv_2}
-\frac{R_k}{1-\beta_k} + \frac{X_k (1 - \gamma_k)}{1 -
(1-\gamma_k)(1-\beta_k)} & = & 0 \\ \nonumber
-\frac{S_k}{1-\gamma_k} + \frac{Y_k}{\gamma_k} + \frac{X_k (1 -
\beta_k)}{1 - (1-\gamma_k)(1-\beta_k)} & = & 0,
\end{eqnarray}
yielding the solution
\begin{equation} \label{eq:initial_solution}
\hat{\beta}_k = \frac{\frac{X_k}{\sum_{i=k}^{n+1} X_i} -
\frac{Y_k}{\sum_{i=k}^{n+1} Y_i}}{1 - \frac{Y_k}{\sum_{i=k}^{n+1}
Y_i}}, \quad \hat{\gamma}_k = \frac{Y_k}{\sum_{i=k}^{n+1} Y_i}.
\end{equation}
One can verify local maximality of $(\hat{\beta}_k, \hat{\gamma}_k)$
from $l_k$'s Hessian. Its global optimality then follows from
$l_k(\beta_k, \gamma_k)$'s continuity and from the fact that it
approaches $-\infty$ near the boundary of its domain's closure.
Eq.~\ref{eq:initial_solution} clearly results in $\hat{\beta}_k <
1$, but there is no guarantee that $\hat{\beta}_k \geq 0$, as its
numerator consists of two terms, each comprising of data from either
the $(+)$ or the $(-)$ channels. As such, they are not constrained
to yield a positive difference, and might result in infeasible
estimates. We then wish to find a \emph{feasible} ML solution, and
we argue that whenever $\hat{\beta}_k < 0$, this solution is
attained at
\begin{equation} \label{eq:constrained_ML}
(\beta_k^{\ast}, \gamma_k^{\ast}) = (0, \frac{X_k +
Y_k}{\sum_{i=k}^{n+1} (X_i + Y_i)}) 
\end{equation}
(see Appendix for justification). This means that whenever we observe a site $k$ for which $\frac{X_k}{\sum_{i=k}^{n+1} X_i} < \frac{Y_k}{\sum_{i=k}^{n+1}
Y_i}$, the best explanation of the observed data is that no
modification occurred at that site and that all $k$-fragments arose
from natural dropoff. Remarkably, this result supports existing approaches to analyzing SHAPE-CE data, whereby sites whose recovered signal is negative are assigned zero reactivity~\cite{Weeks-Review-2010}.

We now allow zero counts when $k \leq n$, while assuming that
$X_{n+1}, Y_{n+1} > 0$. The latter assumption is justified by the
fact that $Y_{n+1} = 0$ is indicative of severe dropoff that could
stem from strong transcription termination at select sites or
reflect a cumulative effect of imperfect transcription elongation
over a long RNA strand. Both situations are avoided by truncating
the analyzed sequence at $n' < n$, such that $Y_{n'+1} > 0$. On the
other hand, $X_{n+1} = 0$ (while $Y_{n+1} > 0$) suggests a ``too
high'' average modification rate, leading to strong signal decay in
the $(+)$ channel. One should then decrease the reagent's
concentration. Nevertheless, zero counts at intermediate sites are
commonly observed in practice. When $X_k = 0$ but $Y_k \neq 0$, it
is straightforward to show that the optimum is determined by
Eq.~\ref{eq:constrained_ML}, whereas the case where $Y_k = 0$ and
$X_k \neq 0$ is optimized by the initial solution in
Eq.~\ref{eq:initial_solution}.

\subsection{Poisson-distributed chemical modification}

In this subsection, we revisit a chemical mapping model that we have
previously developed and used for structure
characterization~\cite{SHAPE-Seq-ML-2011}. The model incorporates an
assumption on the stochastic nature of the underlying chemistry, and
we discuss its implications on ML estimation from SHAPE-Seq data. By
casting the model as a special case of the framework we presented
above we are able to simplify our previous estimation scheme to obtain
\textbf{closed-form ML estimates of relative reactivies for the
  Poisson model in~\cite{SHAPE-Seq-ML-2011}}:
    \begin{equation} \label{eq:relative_reactivity}
     r_k^{\ast} = \max \Big\{ 0, \textstyle \log\big(\textstyle 1 -
     \frac{Y_k}{\sum_{i=k}^{n+1}Y_i}\big) - \log\big(\textstyle 1 -
     \frac{X_k}{\sum_{i=k}^{n+1} X_i}\big)\Big\}.
     \end{equation}

The work in~\cite{SHAPE-Seq-ML-2011} makes an assumption that is
widely used in models of biochemical reactions, whereby the reaction with the modifying reagent follows a Poisson process. Specifically, during
modification, an RNA may be exposed to varying numbers of
electrophile molecules, and we model the number of times it is
exposed to these molecules as a Poisson process of an unknown rate
$c>0$, i.e., we assume that $Prob(\mbox{$i$ exposures}) = \frac{c^i
e^{-c}}{i!}$. It is worth noting that the Poisson framework is
especially suitable for low-incidence settings~\cite{Lang-Poisson-Multinomial-1996},
and that mapping experiments are particularly calibrated to yield
single-hit kinetics, that is, they aim to achieve an average
modification rate of $c \approx 1$. Now, each exposure may result in
the modification of a site, where the site is determined according
to a probability distribution $\Theta=(\theta_1,\ldots,\theta_n)$,
$\sum_{k=1}^n \theta_k = 1$, where $\theta_k$ represents the
\emph{relative reactivity} of site $k$. It is easy to show that the
number of modifications at site $k$ also obeys a Poisson
distribution, with an unknown rate $r_k = c \theta_k$, that is,
$Prob(\mbox{$i$ modifications at site $k$}) = \frac{(c\theta_k)^i
e^{-c\theta_k}}{i!}$. It then follows that $Prob(\mbox{site $k$ is
\emph{not} modified}) = e^{-c\theta_k}$. Setting
\begin{equation} \label{eq:beta_to_theta}
\beta_k := 1 - Prob\left(\mbox{site $k$ is \emph{not}
modified}\right) = 1 - e^{-c\theta_k},
\end{equation}
we can write
\begin{eqnarray}
& & Prob\left(\mbox{first adduct at site $k$}\right) \\ \nonumber &
= & \beta_k \prod_{i=1}^{k-1} (1-\beta_i) = (1 - e^{-c\theta_k})
e^{-c \sum_{i=1}^{k-1} \theta_i}
\\ \nonumber & = & (1 - e^{-c\theta_k}) e^{c
(\sum_{i=k}^{n} \theta_i - 1)} \\ \nonumber & = & e^{c(\sum_{i=k}^n
\theta_i - 1)} - e^{c (\sum_{i=k+1}^n \theta_i - 1)},
\end{eqnarray}
\begin{equation} \label{eq:no_mod_prob}
Prob(\mbox{no modification}) = \prod_{i=1}^{n} (1-\beta_i) = e^{-c}.
\end{equation}
When plugging these expressions into Eq.~\ref{eq:likelihood}, the
likelihood function reduces to that in~\cite{SHAPE-Seq-ML-2011}.
We can therefore use the initial estimates in Eq.~\ref{eq:initial_solution} along with Eq.~\ref{eq:beta_to_theta}
to estimate the distribution $\Theta$ as follows:
\begin{equation} \label{eq:theta_estimate}
\hat{\theta}_k = \frac{1}{\hat{c}} \Big[
\log\big(1 - \frac{Y_k}{\sum_{i=k}^{n+1}
Y_i}\big) - \log\big(1 - \frac{X_k}{\sum_{i=k}^{n+1} X_i}\big)\Big], 
\end{equation}
where the scaling constant $\hat{c}$ is the estimate of the average
modification rate, which is recovered from
Eq.~\ref{eq:no_mod_prob} to equal
\begin{equation} \label{eq:c_estimate}
\hat{c} = - \sum_{i=1}^{n} \log(1 -
\hat{\beta_i}) = \log\Big( \textstyle \frac{Y_{n+1}}{\sum_{i=1}^{n+1} Y_i} \Big) - \log\Big( \textstyle \frac{X_{n+1}}{\sum_{i=1}^{n+1} X_i} \Big).
\end{equation}
It is now apparent that Eqs.~\ref{eq:theta_estimate}
and~\ref{eq:c_estimate} are the outputs of Algorithm 1
in~\cite{SHAPE-Seq-ML-2011}.

When optimization yields negative $\hat{\beta}_k$'s, they correspond to negative $\hat{\theta}_k$'s (see Eq.~\ref{eq:beta_to_theta}). However, when imposing non-negativity, these are projected onto $\beta_k^{\ast} = 0$, and $\theta_k^{\star} \propto \log(1 - \beta_k^{\ast}) = 0$ accordingly.
The revised modification rate estimate now amounts to $c^{\ast}
= - \sum_{i:\hat{\beta_i} > 0} \log(1 -
\hat{\beta_i})$, which is larger than the initial
$\hat{c}$. Consequently, the distribution $\hat{\Theta}$ is updated
such that all negative entries are set to zero, while the others are
effectively scaled down due to the increase from $\hat{c}$ to $c^{\ast}$. In other words, one merely needs to compute the relative reactivity estimate in Eq.~\ref{eq:relative_reactivity}
and then normalize it by $c^{\ast}$ to generate a proper probability distribution $\Theta^{\ast}$. Alternatively, one could apply any other normalization method to the outputs of Eq.~\ref{eq:relative_reactivity}, such as the ones currently used for interpreting SHAPE reactivities~\cite{Vasa-2008, Weeks-Review-2010}. This would retain the relativity between reactivities, while adjusting the dynamic range to a scale that is in line with current settings of subsequent structure prediction modules~\cite{Weeks-Review-2010,Deigan-2008}.

In practice, the formulation of site-specific modifications via $n$ independent Poisson processes, as opposed to the multinomially-distributed choice of a site via $\Theta$, vastly simplifies the likelihood function's derivation and
optimization. In particular, it removes the need for the iterative
likelihood optimization routine that follows Algorithm 1
in~\cite{SHAPE-Seq-ML-2011}. The equivalence between
the two formulations is an instance of general equivalence
between multinomial and Poisson log-linear models with respect to ML
estimation~\cite{Lang-Poisson-Multinomial-1996, Lior-RNA-Seq-Models-Review-2011}.

It is interesting to compare the estimates obtained under a Poisson
assumption with those obtained without it. To qualitatively compare them, consider the relations $\theta_k^{\ast} \propto  -\log (1 - \beta_k^{\ast})$,
and note that $-log(1-x) \approx x$ when $x \approx 0$. This means
that a Poisson assumption is not expected to affect sites with small
reactivity, but on the other hand, it amplifies the estimated
\emph{relative} reactivities at more reactive sites, and more
intensely as the reactivity increases (i.e., as $1 - \beta_k^{\ast}
\rightarrow 0$). It thus exerts its effect by stretching the dynamic
range of reactivities, and thereby might confer more sensitivity to
outliers. Because the effect's intensity depends on
$\beta_k^{\ast}$'s magnitude, it is likely to be more pronounced
under high modification rate conditions, where either the reagent
concentration is high or the RNA entails many highly reactive sites.

\begin{figure*}[htb]
\centering
\includegraphics[width=6.8in]{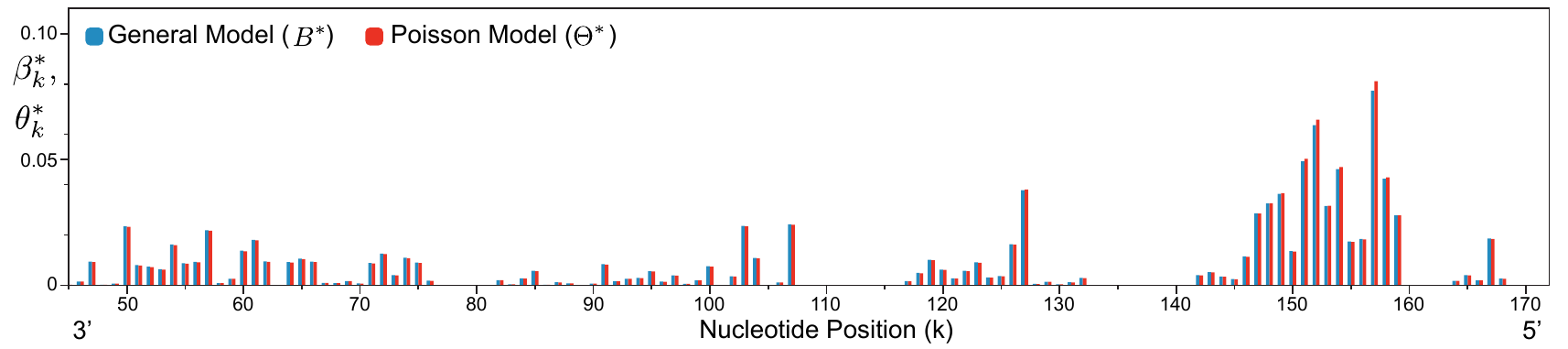}
\caption{Relative reactivity estimates for \emph{S. aureus} plasmid pT181 sense RNA under the general-model and the Poisson-model maximum-likelihood frameworks. Sites 1-45 showed negligible probabilities and were omitted from display.}
\label{fig:Beta_vs_Theta}
\end{figure*}

A quantitative comparison between the two models using experimental
data for the \emph{Staphylococcus aureus} plasmid pT181 sense RNA is
shown in Fig.~\ref{fig:Beta_vs_Theta}. To allow for a fair
comparison between $B^{\ast}$ and $\Theta^{\ast}$, we scaled
$B^{\ast}$ such that its entries sum to $1$. Note that the scaling
factor is larger in the Poisson case, and thus the Poisson-based
reactivities are smaller than their general-model counterparts at
relatively unreactive sites. The data in
Fig.~\ref{fig:Beta_vs_Theta} reveal very mild differences between
the estimates and consequently, minor increase in the dynamic range.
Similar results were observed for a number of other molecules that we
have probed~\cite{SHAPE-Seq-2011}. It is worth noting that the
modification rate in this experiment was estimated at $c^{\ast} =
1.94$ adducts per molecule, which is relatively high and clearly
diverts from single-hit kinetics. This suggests that the Poisson
assumption may not be critical even in the presence of high
modification rate, and that the two model-based schemes may generally
be used interchangeably.

Finally, we stress that the Poisson-based correction aligns more
closely with current CE-based analysis
methodology~\cite{Weeks-Review-2010, Kladwang-SHAPE-errors-2011}, in
the sense that the signals are in fact corrected separately for each
channel and then subtracted, along the lines of Eq.~\ref{eq:relative_reactivity}. In contrast, ML
estimation under the general framework is not amenable to such
decoupling (see Eq.~\ref{eq:initial_solution}). Alongside this seemingly
Poisson-based correction, current analysis guidelines also recommend
using one of two outlier filters~\cite{Weeks-Review-2010}, and these
may remedy the increased sensitivity to outliers that we highlighted
earlier.

\section{Adaptation to quantification by capillary electrophoresis}

Traditional chemical mapping techniques have used electrophoresis
to identify the cDNA fragments and to quantify their
abundances~\cite{Weeks-ChemProbing-Review-2010}. Most recently,
capillary electrophoresis (CE) has been used to detect fluorescently
labeled cDNAs from experiments. CE systems output an electropherogram,
consisting of analog traces that report fluorescence intensity as a function of time. These traces must be extensively processed to  extract quantitative nucleotide information, and while steady progress is being made with the development of computer-aided analysis tools~\cite{Vasa-2008, Laederach-CAFA-2008, Yoon-HiTRACE-2011}, there is still a need for more statistically-robust and automated analysis methods.

Despite the challenges in analyzing CE-based data and the advances
offered by NGS platforms, the conventional approach
is still valuable for two reasons. First, it is currently
cheaper and faster to apply when the multiplexing and sensitivity
advantages of the newer platforms are not needed, and second, it can
be used for probing a few pilot RNAs prior to conducing a
larger-scale NGS-based experiment. These considerations have motivated
us to adapt our SHAPE-Seq ML framework to the case of SHAPE-CE. While
the differences in analog versus digital signal processing are
apparent, an essential difference between SHAPE-Seq and SHAPE-CE data
has to do with the lack of information about complete fragments in CE
settings. This is due to large amounts of long fragments
under single-hit-kinetics conditions, causing detector saturation. Notably, a strong full-length signal also poses difficulty in accurately quantifying the
last stretch of 10-20 nucleotides~\cite{Steen-JACS-2010}, but in this
work, we only address the first issue and assume that all other peaks
are quantifiable. 

\subsection{Maximum likelihood framework}

Here, we show that in the absence of complete-fragment information,
the relative reactivities at sites $1$ to $n-1$ are estimated using
the following formula ($1\leq k \leq n-1$):
\begin{equation} \label{eq:CE_scheme}
\beta_k^{(CE)\ast} = \max \left\{0, \mbox{\ }
\frac{\frac{\tilde{X}_k}{\sum_{i=k}^{n} \tilde{X}_i} -
\frac{\tilde{Y}_k}{\sum_{i=k}^{n} \tilde{Y}_i}}{1 -
\frac{\tilde{Y}_k}{\sum_{i=k}^{n} \tilde{Y}_i}}\right\},
\end{equation}
where $(\tilde{X_1}, \ldots, \tilde{X_n})$ and $(\tilde{Y_1},
\ldots, \tilde{Y_n})$  are the areas under the detected peaks in the $(+)$ and $(-)$ channel traces, respectively. We also show that $\beta_n^{(CE)\ast}$ cannot be determined from the available information.

Our result is possible due to the very recent automation of the trace-alignment and peak-fitting steps with the HiTRACE software~\cite{Yoon-HiTRACE-2011}. This, in turn, generates quantifiable nucleotide reactivity data, in the form of integrated peak areas, prior to signal correction and scaling. The peak areas can then be used in place of the digital sequence-counts to deconvolve the effects of over-modification and natural dropoff on the observed $(+)$ channel signal, as was recently done in~\cite{Kladwang-SHAPE-errors-2011} by means of an optimization routine.
This is in contrast to previous analysis tools, where signal
correction and scaling were needed prior to the application of semi-manual alignment, fitting, and integration routines~\cite{Vasa-2008}.

We assume that the peak areas from the $(+)$ and $(-)$ channels
are proportional to the fragment counts as follows: $\tilde{X_k} =
\delta X_k$ and $\tilde{Y_k} = \epsilon Y_k$ for all $1\leq k \leq
n$ where $\delta$ and
$\epsilon$ are unknown positive constants. The two potentially \emph{different} constants reflect
experimental variation between the channels, including differences
in such factors as molecular concentrations and dye
intensities~\cite{Vasa-2008}. Currently, these are corrected for by
scaling the $(-)$ channel signal by a constant factor that is set
either manually following visual inspection~\cite{Vasa-2008} or
automatically via an optimization routine~\cite{Kladwang-SHAPE-errors-2011}. We will see that in our ML scheme there is no benefit in applying such a ``correction''.

The likelihood of observing the peak areas is given by
\begin{eqnarray} \label{eq:likelihood_CE_1}
\mathcal{L^{CE}}(B, \Gamma) & = & \prod_{k=1}^n
\Big[\gamma_k \prod_{i=1}^{k-1} (1-\gamma_i)\Big]^{\frac{\tilde{Y}_k}{\epsilon}} \\
\nonumber & & \prod_{k=1}^n \Big[\Big( 1 -
(1-\gamma_k)(1-\beta_k)\Big) \\ \nonumber & & \prod_{i=1}^{k-1}
(1-\gamma_i)(1-\beta_i)\Big]^{\frac{\tilde{X}_k}{\delta}},
\end{eqnarray}
and the log-likelihood function is then written as
\begin{eqnarray} \label{eq:log_likelihood_CE}
\log\mathcal{L^{CE}}(B, \Gamma) & = & \sum_{k=1}^{n-1}
l_k^{CE}(\beta_k, \gamma_k) + \frac{1}{\epsilon} \tilde{Y}_n \log \gamma_n \\
\nonumber & & + {} \frac{1}{\delta} \tilde{X}_n \log\big(1 -
(1-\gamma_n)(1-\beta_n)\big),
\end{eqnarray}
where
\begin{eqnarray}
l_k^{CE}(\beta_k, \gamma_k) & = & U_k \log (1-\gamma_k)\\
\nonumber & & + {} V_k \log (1-\beta_k) + \frac{1}{\epsilon}
\tilde{Y}_k \log \gamma_k \\ \nonumber & & + {} \frac{1}{\delta}
\tilde{X}_k \log\big( 1 - (1-\gamma_k)(1-\beta_k)\big),
\end{eqnarray}
and $U_k=\sum_{i=k+1}^{n} (\frac{1}{\delta}\tilde{X}_i +
\frac{1}{\epsilon} \tilde{Y}_i)$, $V_k=\frac{1}{\delta}
\sum_{i=k+1}^{n} \tilde{X}_i$.

Assuming all peak areas are nonzero, we can repeat the derivation in the previous section, while noting two differences: first, different coefficients appear in the equations and second, the last equation (when $k=n$) is
different. Whereas the first difference is minor when all observables are positive, the second one leads to an important difference in the ML solution. Therefore, we start by optimizing $l_k^{CE}(\beta_k, \gamma_k)$ when $1 \leq k \leq n-1$ to obtain
\begin{equation} \label{eq:SHAPE_CE_ML}
\hat{\beta}_k^{CE} = \frac{\frac{\tilde{X}_k}{\sum_{i=k}^{n} \tilde{X}_i} -
\frac{\tilde{Y}_k}{\sum_{i=k}^{n} \tilde{Y}_i}} {1 -
\frac{\tilde{Y}_k}{\sum_{i=k}^{n} \tilde{Y}_i}}, \quad \hat{\gamma}_k^{CE} = \frac{\tilde{Y}_k}{\sum_{i=k}^{n} \tilde{Y}_i}.
\end{equation}
In the special cases where $\tilde{X}_k$ or $\tilde{Y}_k$ are zero,
but $U_k$ and $V_k$ are positive and $U_k > V_k$, we obtain
$\big(\hat{\beta}_k^{CE}, \hat{\gamma}_k^{CE}\big) = \big(0,
\frac{\tilde{Y}_k}{\sum_{i=k}^{n} \tilde{Y}_i}\big)$ and
$\big(\hat{\beta}_k^{CE}, \hat{\gamma}_k^{CE}\big) =
\big(\frac{\tilde{X}_k}{\sum_{i=k}^{n} \tilde{X}_i}, 0\big)$,
respectively. It can also be verified that these points correspond
to a global maximum, independently of the $\delta$ and $\epsilon$
values. When $\hat{\beta}_k^{CE} < 0$, one can repeat previous
arguments to show that the constrained maximum is attained at
$\big(\beta_k^{(CE)\ast}, \gamma_k^{(CE)\ast}\big) = \big(0,
\frac{\frac{1}{\delta} \tilde{X}_k + \frac{1}{\epsilon}
\tilde{Y}_k}{U_{k-1}}\big)$. Taken together, these results lead to the formulation in Eq.~\ref{eq:CE_scheme}. Importantly, one cannot evaluate $\gamma_k^{(CE)\ast}$ without knowledge of the relative scaling factor $\frac{\delta}{\epsilon}$, however, our goal is to estimate the relative reactivities, and these are independent of $\frac{\delta}{\epsilon}$.

While the estimates in Eq.~\ref{eq:CE_scheme} bear great similarity to the NGS-based estimates, the case where $k=n$ reveals a different scenario, as the missing $\tilde{X}_{n+1}$ and $\tilde{Y}_{n+1}$ hamper $\beta_n$'s
estimation. In this case, we optimize the function $l_n^{CE}(\beta_n, \gamma_n) = \frac{1}{\epsilon} \tilde{Y}_n \log \gamma_n + \frac{1}{\delta} \tilde{X}_n \log\big(1 - (1-\gamma_n)(1-\beta_n)\big)$, which is maximized at $\hat{\gamma}_n = 1$, where its value is independent of $\hat{\beta}_n$'s value.
Consequently, one cannot determine $\hat{\beta}_n$. Moreover, one cannot recover the exact fraction of modified molecules, as it depends on all $n$ reactivities as follows:
\begin{equation} \label{eq:frac_modified}
f_{mod} = Prob(\mbox{molecule is modified}) = 1 -
\prod_{i=1}^{n} (1-\beta_i).
\end{equation}
A possible way to circumvent this limitation is by excluding site $n$ from analysis and evaluating only $\beta^{(CE)\ast}_1, \ldots , \beta^{(CE)\ast}_{n-1}$ based on all available peak areas, including $\tilde{X}_n$ and $\tilde{Y}_n$. In practice, the effect of such omission is likely to be minor, since the studied RNA sequence is typically embedded in between auxiliary RNA constructs, called structure  cassettes~\cite{Weeks-Nature-Protocols-2006}, and site $n$ is therefore included in one of these cassettes \cite{SHAPE-Seq-2011}. We can then approximate the modification fraction by $\hat{f}_{mod}^{CE} \approx 1 - \prod_{i=1}^{n-1} \big(1-\beta^{(CE)\ast}_i\big)$, where the goodness of this
approximation depends on how large $\tilde{X}_n$ is in comparison to
the missing $\tilde{X}_{n+1}$. This is because the approximation
implicitly assumes that $\beta^{(CE)\ast}_n = 0$, whereas in the
presence of a full-length signal we would have $\beta^{(CE)\ast}_n =
1 - \frac{\tilde{X}_{n+1}}{\tilde{X}_{n}+\tilde{X}_{n+1}}\times
\frac{\tilde{Y}_{n}+\tilde{Y}_{n+1}}{\tilde{Y}_{n+1}}$, which might
diverge from zero whenever
$\frac{\tilde{X}_{n+1}}{\tilde{X}_{n}+\tilde{X}_{n+1}}$ diverges
from $1$. The reasoning behind setting $\hat{\beta}_n = 0$ is
twofold: first, our past experience with analyzing SHAPE-Seq data
shows that the structure cassettes tend to display negligible
reactivities, and second, the observed counts $X_{n+1}$ and
$Y_{n+1}$ were very large compared to \emph{any} other count. For
example, in mapping experiments we conducted, $X_{n+1}$ amounted to
approximately $10\%$ of the total $(+)$-channel reads and $Y_{n+1}$
amounted to $15\%-20\%$ of the total $(-)$-channel reads
\cite{SHAPE-Seq-ML-2011, SHAPE-Seq-2011}, whereas the rest of the
reads were associated with a total of $100-200$ nucleotides. Based
on these observations and on the premise that SHAPE-CE statistics
should follow similar patterns, we believe that this approximation
is fairly accurate in general. Nonetheless, one must keep in mind
that this may not always be the case, especially when studying long
RNAs, where cumulative dropoff effects result in severe signal
attenuation and consequently relatively weak full-length signal. In
addition, we point out another subtle difference between the
scheme's implementations under the two platforms. Specifically, we
may not assume that the CE-based $U_k$ and $V_k$ are always
positive, whereas we made a similar assumption when we derived
NGS-based estimates. That assumption was based on the fact
that probing experiments can be designed such that a strong
full-length signal arises. While this also applies to CE-based
protocols, the absence of a full-length signal might complicate
analysis whenever $\tilde{X}_n = 0$ or $\tilde{Y}_n = 0$. However, this can be
easily remedied by converting these zeros into very small constants,
such that the resulting approximations are negligible.

To summarize, building on recent contributions to the automation of
CE-based analog signal processing~\cite{Yoon-HiTRACE-2011}, our
method facilitates a simple and completely automated data analysis
pipeline for CE-based chemical mapping probes, and in particular,
for SHAPE-CE. Another interesting point arising from our derivation is
that our ML scheme is invariant under background-signal scaling.

\subsection{Effects of full-length signal information on ML estimation}

Our analysis highlights the fact that the difference in ML estimation
between the two platforms lies essentially in the presence or
absence of a full-length signal (compare Eqs.~\ref{eq:Seq_ML_scheme}
and~\ref{eq:CE_scheme}). In this subsection, we first use our derivation to qualitatively explore the potential impact of this difference. We then quantify its effects by deleting the full-length signal information from SHAPE-Seq data to mimic SHAPE-CE data, such that the estimates under both platforms can be compared.

Before we start, we stress that this difference between platforms pertains only to RNAs that are no longer than 400-600 nucleotides~\cite{Weeks-ChemProbing-Review-2010}, a limitation
imposed by RT's imperfect processivity, as well as to RNAs that do
not contain major transcription barriers that result in severe dropoff. RNAs of these two types are probed by annealing multiple primers at various sites~\cite{Deigan-2008, Weeks-HIV-Plos-Bio-2008, Weeks-HIV-Nature-2009}, in which case a full-length signal is not obtained from most primer locations even when the fragments are
sequenced. In such settings, NGS and CE platforms generate similar information, and analysis should follow the lines of the CE framework. In what follows, we simplify the exposition by using the same notation for $X_k$ and $\tilde{X}_k$, and similarly for $Y_k$ and $\tilde{Y}_k$.

\begin{figure*}[!t]%
\centering
\subfigure[]{\includegraphics[width=6.8in]{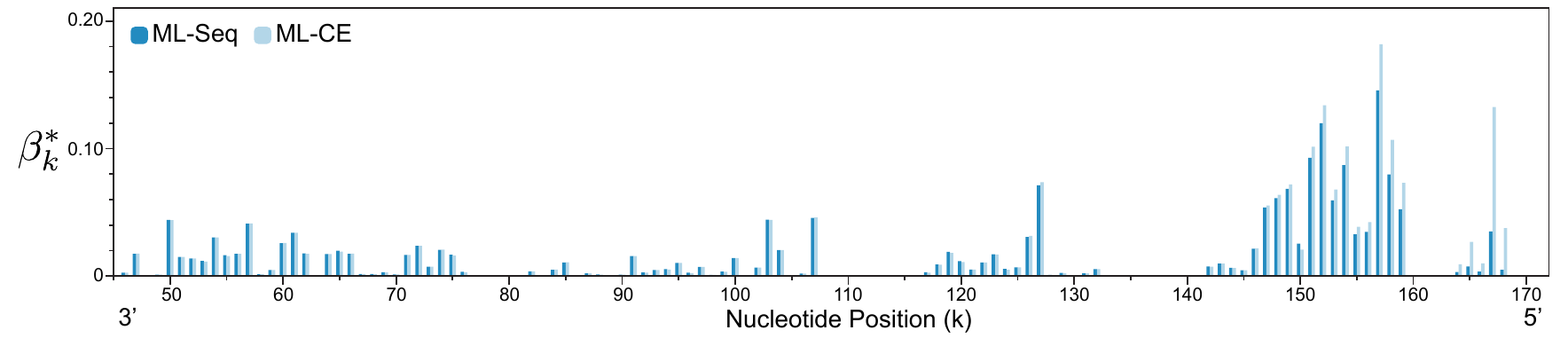}%
\label{fig:Seq_vs_CE_pT181}}\\
\subfigure[]{\includegraphics[width=6.8in]{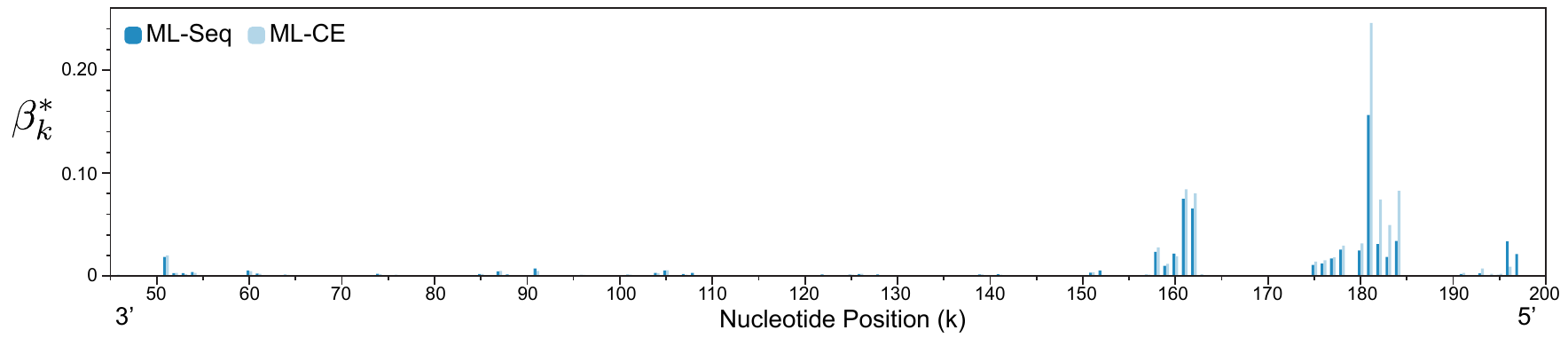}%
\label{fig:Seq_vs_CE_RNaseP}} \caption{Reactivity estimates
in the presence and absence of complete-fragment information for \emph{S. aureus} plasmid pT181 sense RNA (a) and \emph{B. subtilis} RNase P RNA (b). Sites at the $3'$ end that  showed negligible reactivities were omitted from display.}
\label{fig:Seq_vs_CE}
\end{figure*}

To better understand the implications of not recording the
full-length information, we rewrite the initial estimates as
\begin{equation} \label{eq:one_minus_beta}
1-\hat{\beta}^{SEQ}_k = \frac{1 - \frac{X_k}{\sum_{\boldsymbol i = \boldsymbol k}^{\boldsymbol n + \boldsymbol 1} X_i}}{1 - \frac{Y_k}{\sum_{\boldsymbol i = \boldsymbol k}^{\boldsymbol n+ \boldsymbol 1} Y_i}}, \mbox{\ \ \ } 1-\hat{\beta}^{CE}_k = \frac{1 - \frac{X_k}{\sum_{\boldsymbol i = \boldsymbol k}^{\boldsymbol n} X_i}} {1 - \frac{Y_k}{\sum_{\boldsymbol i = \boldsymbol k}^{\boldsymbol n} Y_i}},
\end{equation}
and consider the following two examples.
\begin{enumerate}
\item Assume for simplicity that the $(+)$ and $(-)$ RNA pools are
the same size, i.e., $\sum_{i=1}^{n+1} X_i = \sum_{i=1}^{n+1}
Y_i$, and suppose we probe a highly reactive molecule, or
alternatively, use high reagent concentration. Both scenarios divert from single-hit kinetics toward higher modification rates, resulting in large dropoff in the $(+)$ channel. As an extreme case, assume that the proportion of complete fragments in the $(+)$ channel is negligible, such that $\sum_{i=1}^{n} X_i
\approx \sum_{i=1}^{n+1} X_i$, while it is significant in the $(-)$
channel. In this case, the difference between $1-\hat{\beta}^{SEQ}_k$ and $1-\hat{\beta}^{CE}_k $ amounts to the difference between $\frac{Y_k}{\sum_{i=k}^{n+1}Y_i}$ and $\frac{Y_k}{\sum_{i=k}^{n} Y_i}$. Because $Y_{n+1}$ occupies a major fraction of the $(-)$ channel pool, the denominator of the right-hand expression is larger than the left-hand one, and consequently
$\hat{\beta}^{SEQ}_k > \hat{\beta}^{CE}_k$. This example illustrates
that the missing information might lead to different estimates, and
that under some scenarios, it results in under-estimation of all
reactivities. Furthermore, the effect becomes more pronounced as we
progress toward site $n$, since $Y_{n+1}$ occupies increasingly
larger fractions of $\sum_{i=k}^{n+1} Y_i$. For this reason, the
\emph{relative} reactivities are distorted as well, even when all the
$\beta_k$'s are under-estimated.

\item In this example, we still assume that $\sum_{i=1}^{n+1} X_i =
\sum_{i=1}^{n+1} Y_i$, but we now require both $X_{n+1}$ and $Y_{n+1}$
to represent significant fractions of the overall pools. We further assume that at the last two sites we observe $Y_n = Y_{n-1} = 3$, $X_{n-1} = 30$, and $X_n =
6$. Then, for site $k = n-1$ we obtain $\hat{\beta}^{CE}_{n-1} = 1 -
(1-\frac{30}{36}) / (1-\frac{3}{6}) = \frac{2}{3}$, but on the other
hand, when $X_{n+1}, Y_{n+1}$ are large enough (e.g., on the order
of thousands), we have $\hat{\beta}^{SEQ}_{n-1} \approx 0$. Here,
unknown large end-signals lead to the misinterpretation of minor $(+)$ channel signals (or merely system noise) as indicative of high reactivity. This stems from misrepresentation of the molecular pool composition by $X_1,\ldots , X_n$, and as before, the effect tends to intensify as we get closer to site $n$.
\end{enumerate}

These examples describe very specific and perhaps extreme
scenarios, and clearly, it is difficult to predict the effect of a
given pair $X_{n+1}, Y_{n+1}$ on the estimates of the entire RNA
sequence, as these also depend on the underlying fragment-length
distributions. While distortion may certainly be small at many sites, such differences accumulate when all estimates are jointly aggregated into an estimate of the modification fraction $f_{mod}$. To demonstrate this cumulative effect, we first assume that the initial estimate $\hat{B}$ consists entirely of nonnegative $\hat{\beta}_k$'s under both platforms, and so no zeroing is applied. It is then easy to see that
\begin{eqnarray}
\hat{f}_{mod}^{CE} & \approx & 1 - \frac{X_n}{Y_n} \times
\frac{\sum_{i=1}^{n} Y_i}{\sum_{i=1}^{n} X_i} \\ \nonumber
\hat{f}_{mod}^{SEQ} & = & 1 - \frac{X_{n+1}}{Y_{n+1}} \times
\frac{\sum_{i=1}^{n+1} Y_i}{\sum_{i=1}^{n+1} X_i},
\end{eqnarray}
where the inequality is due to an unknown $\hat{\beta}_n$ under CE settings (see discussion following Eq.~\ref{eq:frac_modified}). For simplicity, we again assume equally sized $(+)$ and $(-)$ RNA pools, in which case $\frac{X_{n+1}}{Y_{n+1}}$ completely determines $\hat{f}_{mod}^{SEQ}$, whereas $\hat{f}_{mod}^{CE}$ is affected by site $n$'s data ratio, $\frac{X_n}{Y_n}$, as well as by $X_{n+1}$ and $Y_{n+1}$'s portions of the entire RNA pools (rather than by their ratio). It thus appears that the CE-based estimate is more susceptible to the actual experimental conditions and to noise toward the signal's end. Yet, in practice, many of the point estimates are zeroed out and so the formulation above may not accurately capture the true estimates.

We conclude with two examples computed from SHAPE-Seq data, where we
compare the NGS- and CE-based ML frameworks for the
\emph{Staphylococcus aureus} pT181 RNA and for the \emph{Bacillus
subtilis} RNase P RNA specificity domain. The CE case was analyzed
by deleting the $(+)$ and $(-)$ end signals, thus reflecting our
assumption that SHAPE-CE data closely resembles SHAPE-Seq data, as
we have previously observed for these two
RNAs~\cite{SHAPE-Seq-2011}. The estimated reactivities shown in
Fig.~\ref{fig:Seq_vs_CE} support our observations, and a clear trend
of divergence between the estimates under the two frameworks is
apparent toward the $5'$ end, where in these two cases the CE data
result in over-estimation. Interestingly, the divergence in the
fraction of modified molecules was minor in the pT181 case
($\hat{f}_{mod}^{CE} \approx 0.9$ vs. $\hat{f}_{mod}^{SEQ} \approx
0.86$) and amounted to approximately $20\%$ for RNase P
($\hat{f}_{mod}^{CE} \approx 0.62$ vs. $\hat{f}_{mod}^{SEQ} \approx
0.52$). Our results thus point out to a potential shortcoming of
likelihood-based signal recovery schemes, when used in conjunction
with CE systems. It is important to stress, however, that the
previous and less automated method developed by the Weeks and
Giddings labs~\cite{Vasa-2008} does not suffer from this
shortcoming. This is because it relies on visual assessment of the
signal's decay and its subsequent correction, whereas
likelihood-based methods such as ours and the one reported
in~\cite{Kladwang-SHAPE-errors-2011} explicitly utilize the observed
frequencies to correct the signal. At the same time, relying on user
feedback poses challenges to the reproducibility and accuracy of
analysis, as discussed in~\cite{SHAPE-Seq-ML-2011}.

\section{Conclusion}

In this work, we presented a model and a maximum-likelihood framework, which lead to simple closed-form reactivity estimates, and which are applicable to chemical probes that use either CE or NGS for transcript quantification. We used this general framework to directly compare the estimates obtained with the two detection platforms, and concluded that lack of full-length signal information in CE settings degrades the estimates quality, and hence SHAPE-Seq is a more informative, and potentially more accurate, technique. Yet, it remains to determine the effects of this missing information on structure prediction's accuracy in order to clearly characterize the benefits of using new protocols such as SHAPE-Seq. 

\appendix

To justify our claim that Eq.~\ref{eq:constrained_ML} pertains to the feasible ML solution, we first assume that $\beta_k = 0$ and then optimize $l_k(0, \gamma_k) = S_k \log (1-\gamma_k) + (X_k + Y_k) \log \gamma_k$ to obtain $\gamma_k^{\ast}$ as its maximizing argument. Hence, this point represents the maximum over all points on one edge of the constrained optimization domain $\mathcal{D} = \{(\beta_k, \gamma_k) : 0\leq \beta_k \leq 1, \mbox{\ } 0 \leq
\gamma_k \leq 1\}$. Now, assume that our claim is not correct, i.e.,
the maximum over $\mathcal{D}$ is not attained at a point where
$\beta_k = 0$. Then, there exists a point $(\tilde{\beta}_k,
\tilde{\gamma}_k)$ such that $l_k(\tilde{\beta}_k, \tilde{\gamma}_k)
> l_k(\beta_k^{\ast}, \gamma_k^{\ast}) \geq l_k(0, \gamma_k)$ for
any $0 < \gamma_k < 1$. Clearly, $(\tilde{\beta_k},
\tilde{\gamma_k})$ must lie in $\mathcal{D}$'s interior since $l_k$
approaches $-\infty$ near all other three edges of $\mathcal{D}$.
Next, we construct a rectangular compact set $\mathcal{E} =
\{(\beta_k, \gamma_k) : 0 \leq \beta_k \leq a < 1, \mbox{\ } 0 < b
\leq \gamma_k \leq c < 1\} \subset \mathcal{D}$ around
$(\tilde{\beta}_k, \tilde{\gamma}_k)$, where we choose $a, b$ and
$c$ such that  $l_k(\tilde{\beta}_k, \tilde{\gamma}_k)$ exceeds
$l_k$'s values over  $\mathcal{E}$'s boundary. This is possible due
to the function's decline to $-\infty$ as $\beta_k$ approaches $1$
and as $\gamma_k$ approaches $\{0,1\}$ and because
$l_k(\tilde{\beta}_k, \tilde{\gamma}_k)$ is greater than $l_k$'s
values at the fourth edge (where $\beta_k = 0$). Since $\mathcal{E}$
is compact, $l_k$ must attain a global maximum over it, but it
follows from the construction that the maximum must lie in its
interior. This, in turn, means that this maximum must also be a
stationary point (since $l_k$ is differentiable over $\mathcal{E}$),
which contradicts the fact that the only stationary point is
$(\hat{\beta}_k, \hat{\gamma}_k)$, which lies outside of
$\mathcal{D}$.


\section*{Acknowledgment}
We thank Rhiju Das and Adam Siepel for comments and insights on our
previous work that inspired us to formulate the general model
presented in this manuscript.




\begin{thebibliography}{10}

\bibitem{Sharp-RNA-Centrality-Review-2009} P.~A. Sharp, ``The centrality of RNA,'' \emph{Cell}, vol. 136, pp. 577-580, Feb. 2009.

\bibitem{Chang-lncRNA-Review-2009} O. Wapinski and H.~Y. Chang, ``Long noncoding RNAs and human disease,'' \emph{Trends Cell Biol.}, vol. 21, pp. 354-361, June  2011.

\bibitem{Collins-RNA-Syn-Bio-Review-2006} F.~J. Isaacs, D.~J. Dwyer, and J.~J. Collins, ``RNA synthetic biology,'' \emph{Nat. Biotechnol.}, vol. 24, pp. 545-554, May 2006.

\bibitem{Weeks-ChemProbing-Review-2010} K.~M. Weeks, ``Advances in RNA structure analysis by chemical probing,'' \emph{Curr. Op. Struct. Biol.}, vol. 20, pp. 295-304, June 2010.

\bibitem{Rocca-Serra-Mapping-Exp-Standards-2011} P. Rocca-Serra et al., ``Sharing and archiving nucleic acid structure mapping data,'' \emph{RNA}, vol. 17, pp. 1204-1212, June 2011.

\bibitem{Mathews-2004} D.~H. Mathews, M.~D. Disney, J.~L. Childs, S.~J. Schroeder, M. Zuker, and D.~H. Turner, ``Incorporating chemical modification constraints into a dynamic programming algorithm for prediction of RNA secondary structure,'' \emph{Proc. Natl. Acad. Sci. USA}, vol. 101, pp. 7287-7292, May 2004.

\bibitem{Weeks-Review-2010} J.~T. Low and K.~M. Weeks, ``SHAPE-directed RNA secondary structure prediction,'' \emph{Methods}, vol. 52, pp. 150-158, Oct. 2010.

\bibitem{Weeks-Nature-Protocols-2006} K.~A. Wilkinson, E.~J. Merino, and K.~M. Weeks, ``Selective 2'-hydroxyl acylation analyzed by primer extension
(SHAPE): quantitative RNA structure analysis at single nucleotide
resolution,'' \emph{Nat. Protoc.}, vol. 1, pp. 1610-1616, Nov. 2006.

\bibitem{SHAPE-Seq-2011} J.~B. Lucks et al., ``Multiplexed RNA structure characterization with selective 2'-hydroxyl acylation analyzed by primer extension sequencing (SHAPE-Seq),'' \emph{Proc. Natl. Acad. Sci. USA}, in press.

\bibitem{SHAPE-Seq-ML-2011} S. Aviran et al., ``Modeling and automation of sequencing-based characterization of RNA structure,'' \emph{Proc.
Natl. Acad. Sci. USA}, in press.

\bibitem{Lang-Poisson-Multinomial-1996} J.~B. Lang, ``On the comparison of multinomial and Poisson log-linear models,'' \emph{J. Royal Stat. Soc. B} 58:253-266, Jan. 1996.

\bibitem{Vasa-2008} S.~M. Vasa, N. Guex, K.~A. Wilkinson, K.~M. Weeks, and M.~C. Giddings, ``ShapeFinder: A software system for high-throughput quantitative analysis of nucleic acid reactivity information resolved by capillary electrophoresis,'' \emph{RNA}, vol. 14, pp. 1979–1990, Oct. 2008.

\bibitem{Deigan-2008} K.~E. Deigan, T.~W. Li, D.~H. Mathews, and K.~M. Weeks, ``Accurate SHAPE-directed RNA structure determination,'' \emph{Proc. Natl.
Acad. Sci. USA}, vol. 106, pp. 97-102, Dec. 2008.

\bibitem{Lior-RNA-Seq-Models-Review-2011} L. Pachter, ``Models for transcript quantification from RNA-Seq,'' arXiv:1104.3889v2 [q-bio.GN].


\bibitem{Kladwang-SHAPE-errors-2011} W. Kladwang W, C.~C. VanLang, P. Cordero, and R. Das, ``Understanding the errors of SHAPE-directed RNA structure modeling,'' arXiv:1103.5458v2 [q-bio.QM].

\bibitem{Laederach-CAFA-2008} S. Mitra, I.~V. Shcherbakova, R.~B. Altman, M. Brenowitz, and A. Laederach, ``High-throughput single-nucleotide structural mapping by capillary automated footprinting analysis,'' \emph{Nucl. Acid Res.}, vol. 36, pp. e63:1-10, May 2008.

\bibitem{Yoon-HiTRACE-2011} S. Yoon et al., ``HiTRACE: High-throughput robust analysis for capillary electrophoresis,'' \emph{Bioinformatics}, vol. 27, pp. 1798–1805, June 2011.

\bibitem{Steen-JACS-2010} K.~A. Steen., A. Malhotra A, and K.~M. Weeks, ``Selective 2'-hydroxyl acylation analyzed by protection from exoribonuclease,'' \emph{J. Am. Chem. Soc.}, vol. 132, pp. 9940-9943, July 2010.

\bibitem{Weeks-HIV-Plos-Bio-2008} K.~A. Wilkinson et al., ``High-throughput SHAPE analysis reveals structures in HIV-1 genomic RNA strongly conserved across distinct biological states,'' \emph{PLoS Biol.}, vol. 6, pp. e96:883-899, Apr. 2008.

\bibitem{Weeks-HIV-Nature-2009} J.~M. Watts et al., ``Architecture and secondary structure of an entire HIV-1 RNA genome,'' \emph{Nature}, vol. 460, pp. 711-716, Aug. 2009.


\end{thebibliography}
%

\end{document}